\documentclass[aps,prd,showpacs,showkeys,superscriptaddress,twocolumn]{revtex4-1}
\usepackage[colorlinks,linkcolor=blue,anchorcolor=blue,citecolor=blue,urlcolor=blue,breaklinks=true]{hyperref}
\usepackage{graphicx}
\usepackage{amsmath}
\usepackage{amssymb}
\usepackage{slashed}
\usepackage{latexsym}
\usepackage{epsfig}
\usepackage{amsbsy}
\usepackage{array}
\usepackage{changes}
\usepackage{amssymb}
\usepackage{setspace}
\usepackage{bm}
\usepackage{lipsum}
\usepackage{mathrsfs}
\usepackage{float}
\usepackage{color}
\usepackage{graphicx}
\usepackage{subfigure}
\usepackage[T1]{fontenc}
\usepackage{mathptmx}
\DeclareMathAlphabet{\mathcal}{OMS}{cmsy}{m}{n}
\DeclareSymbolFont{largesymbols}{OMX}{cmex}{m}{n}

\begin{document}
\author{Zheng Zhang}
\email{jozhzhang@163.com}
\affiliation{Department of physics, Nanjing University, Nanjing 210093, China}
\author{Chao Shi}
\email{cshi@nuaa.edu.cn}
\affiliation{Department of Nuclear Science and Technology,
Nanjing University of Aeronautics and Astronautics, Nanjing 210016, China}
\author{Xiaofeng Luo}
\email{xfluo@mail.ccnu.edu.cn}
\affiliation{Key Laboratory of Quark and Lepton Physics (MOE) and Institute of Particle Physics, Central China Normal University, Wuhan 430079, China}
\author{Hong-Shi Zong}
\email{zonghs@nju.edu.cn}
\affiliation{Department of physics, Nanjing University, Nanjing 210093, China}
\affiliation{Nanjing Proton Source Research and Design Center, Nanjing 210093, China}
\affiliation{Department of physics, Anhui Normal University, Wuhu, Anhui 241000, China }
\date{\today}

\title{Chiral phase transition in a rotating sphere}

\begin{abstract}
{We study the chiral phase transition of the two-flavor Nambu-Jona-Lasinio (NJL) model in a rotating sphere, which includes both rotation and finite size effects. We find that rotation leads to a suppression of the chiral condensate at a finite temperature, while its effects are smaller than the finite size effects. Our work can be helpful to study the effects relevant to rotation in heavy-ion collisions in a more realistic way.}
\bigskip


\end{abstract}

\maketitle


\maketitle

\section{Introduction}
Exploring the properties of strongly interacting matter under rotation is of particular importance and has attracted extensive attention.  For example, the neutron star, which consists of dense nuclear matter, can rotate rapidly \cite{starreview}. The most vortical strongly interacting fluid with the vorticity about $10^{22} s^{-1}$ has been created in noncentral high energy heavy-ion collision (HIC) experiments~\cite{vorticalfluid}. The properties of rotating strongly interacting matter were also studied with lattice QCD \cite{rotaionla}. It is predicted that rotation can lead to many interesting transport phenomena, such as the chiral vortical effect \cite{chiralv1,chiralv2,chiralv3} and chiral vortical wave \cite{chiralvw}. These phenomena can lead to measurable experimental signals in heavy-ion collisions. Besides the transport properties, rotation can influence the phase structure and phase transition of matter. It is found that rotation will suppress the scalar pairing states \cite{YinJiang}. In the noncentral HIC experiments, it would be very interesting to study how the chiral transition of the hot and dense QCD matter be influenced by such large vorticity. The system of interacting fermions under rotation has been studied in unbounded \cite{unbound1,YinJiang} and bounded \cite{bound1,cylinder} geometries in effective field models, as well as in the holographic approaches \cite{holo1,holo2,holo3}.  However, to give a realistic simulation of the hot and dense QCD matter produced in HIC experiments, one should consider the finite size effects since the typical size of the systems are estimated to be only about 2-10 fm \cite{Palhares2011}. Finite size effects can modify the phase structure of strong interaction and affect the dynamics of phase conversion (see a review in, e.g., \cite{review}). 

In order to consider both the rotation and finite size effects to simulate more realistic conditions such as that in noncentral HIC experiments, we perform our studies for the strongly interacting matter in a rotating sphere with the Nambu-Jona-Lasinio (NJL) model. For simplicity, we consider a rigidly rotating system. For such a system, the direction transverse to the rotation axis must be bounded to make sure the rotating speed cannot exceed the speed of light; otherwise the causality will be violated and some pathologies will occur \cite{pathology2}. To bound the system in a finite region, one should choose a boundary condition. However, different boundary conditions will lead to different physical results \cite{rcylinder}. In a recent work \cite{zzhang}, we compared the chiral phase transition in a sphere (with MIT boundary condition) and in a box (with antiperiodic boundary condition),  and we found the MIT boundary condition induces stronger finite size effects than the antiperiodic boundary condition. We also argued that the MIT boundary condition may be more suitable than the (anti)periodic boundary condition when studying finite size effects due to its confinement feature. Thus, in this paper, we will use the MIT boundary condition to bound the system in a sphere as we did in \cite{zzhang}. 

First, we would like to give a qualitative explanation what the effects of finite size and rotation are. It is well known that, in principle, the spontaneous symmetry breaking only occurs in infinitely large systems~\cite{Weinberg:1996kr}. So, in the NJL model, we expect that the finite size will lead to the restoration of chiral symmetry and thus a lower effective mass \cite{qingwu,zzhang}. For rotation, the global angular momentum will induce a rotational polarization effect which "force" microscopic angular momentum to be parallel to the global angular momentum, so the chiral condensate state which has zero angular momentum will be suppressed~\cite{YinJiang}; thus, the effective mass becomes smaller. Our calculation below will confirm our intuitive understanding.

This paper is organized as follows: In Sec. \ref{II}, we derive the spectrum of free fermions in a rotating sphere. The chiral phase transition of the NJL model in a rotation sphere is studied in Sec. \ref{III} and a summary is given in Sec. \ref{IV}. 

\begin{figure*}[htbp]
\centering

\subfigure[ R=0.985 fm]{
\begin{minipage}{0.43\textwidth}
\centering
\includegraphics[width=1\linewidth]{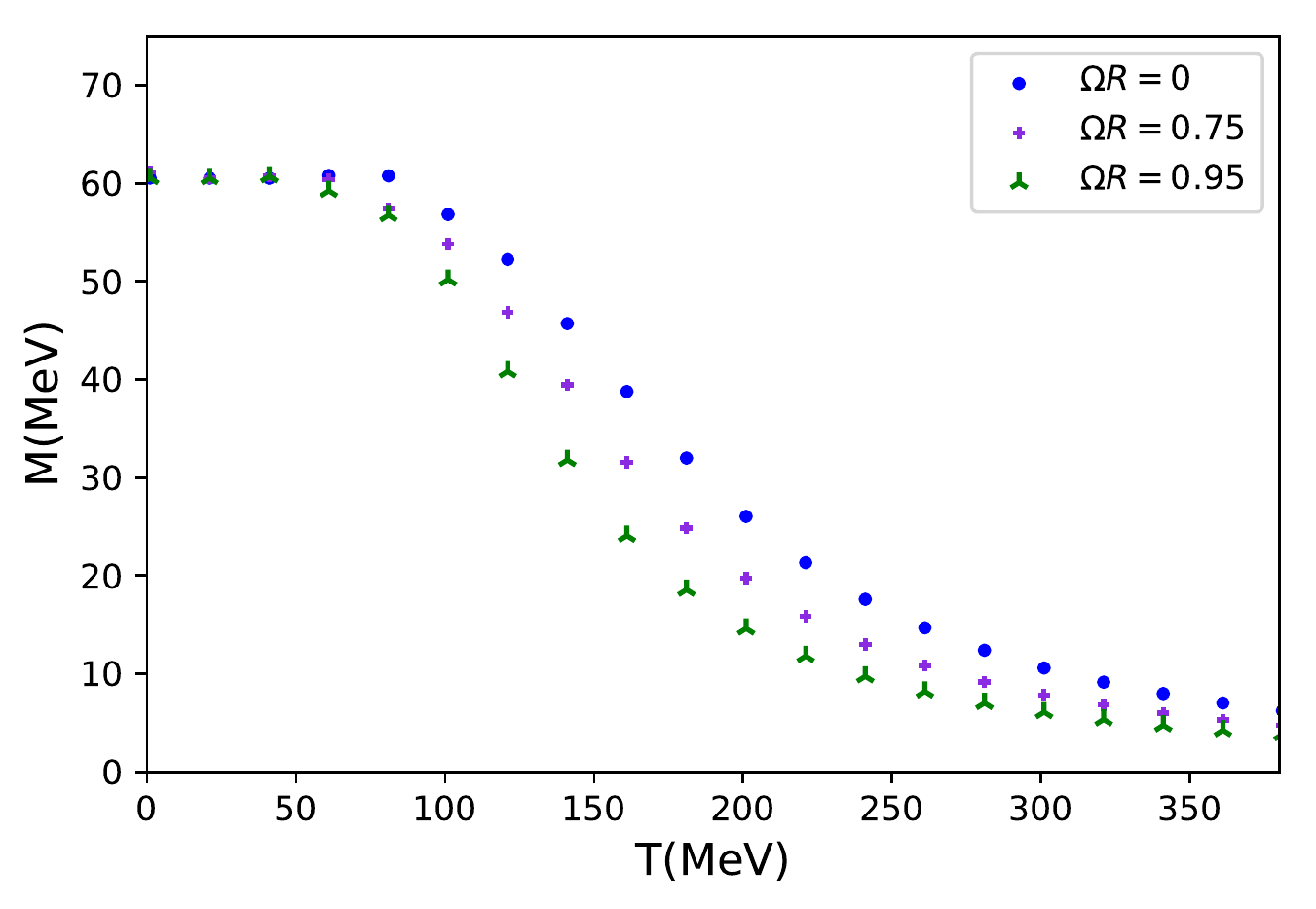} 
\end{minipage}
}
\subfigure[ R=1.576 fm]{
\begin{minipage}{0.43\textwidth}
\centering
\includegraphics[width=1\linewidth]{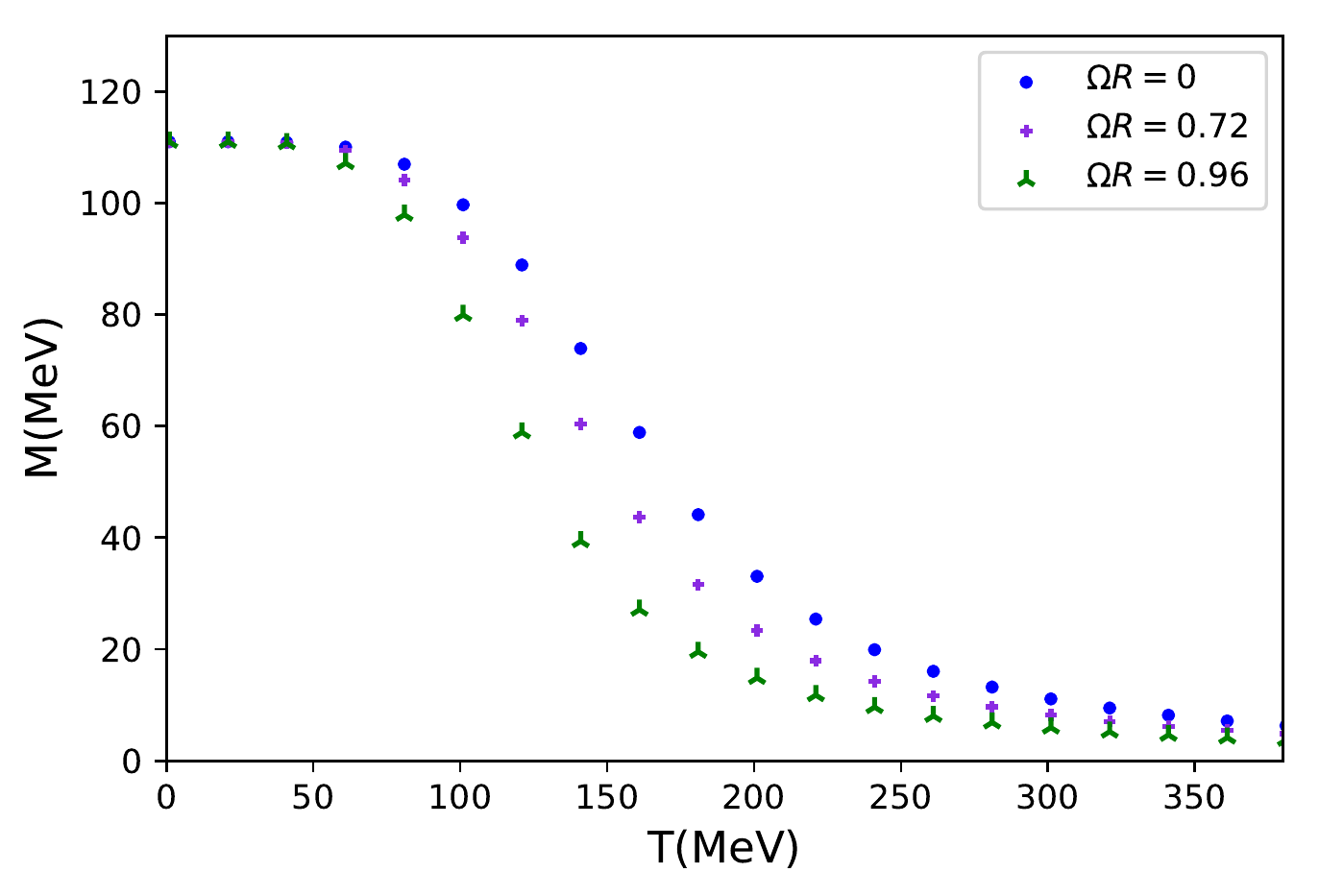} 
\end{minipage}
}
\subfigure[ R=2.955 fm]{
\begin{minipage}[t]{0.43\textwidth}
\centering
\includegraphics[width=1\linewidth]{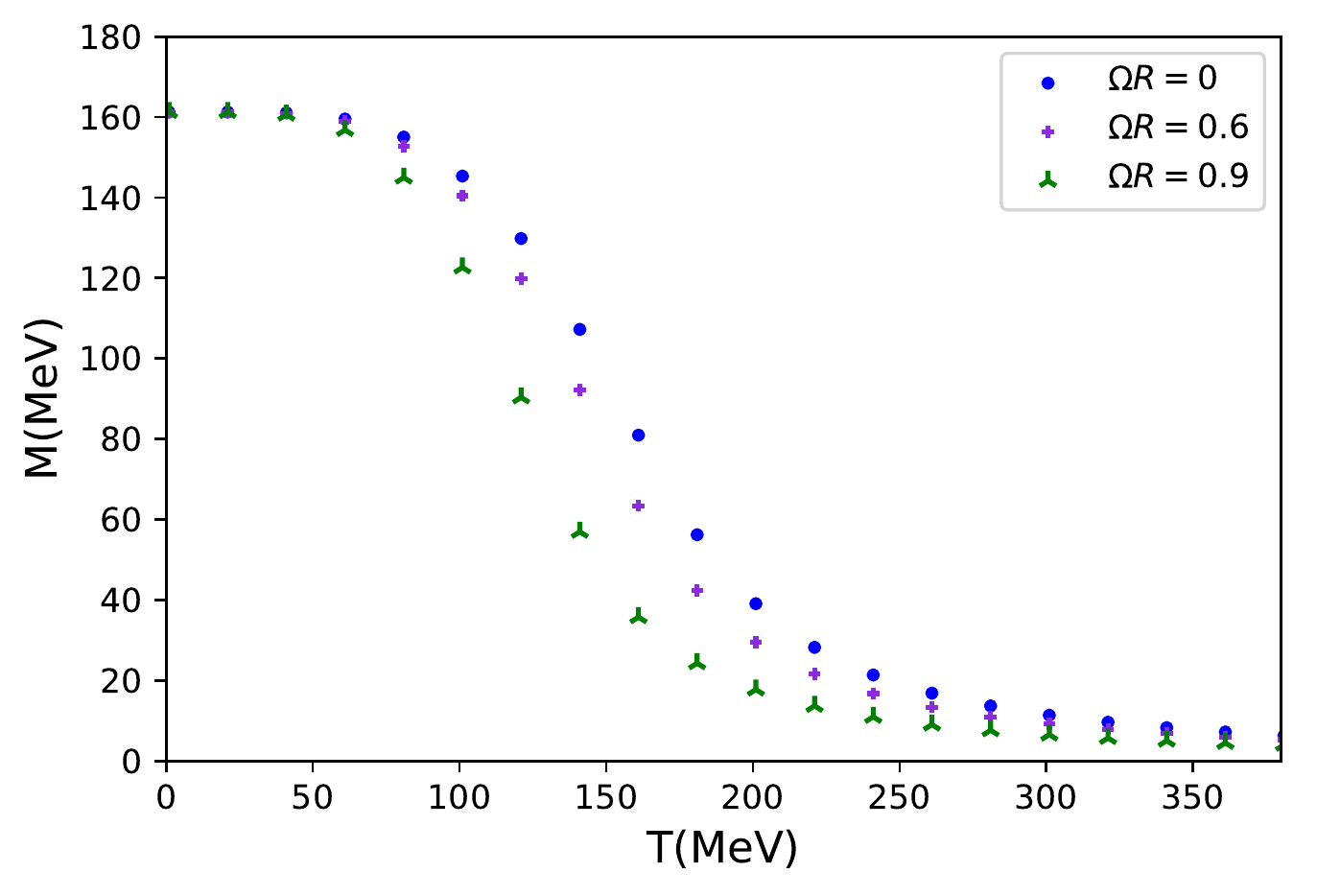} 
\end{minipage}
}
\subfigure[ R=5.91 fm]{
\begin{minipage}[t]{0.43\textwidth}
\centering
\includegraphics[width=1\linewidth]{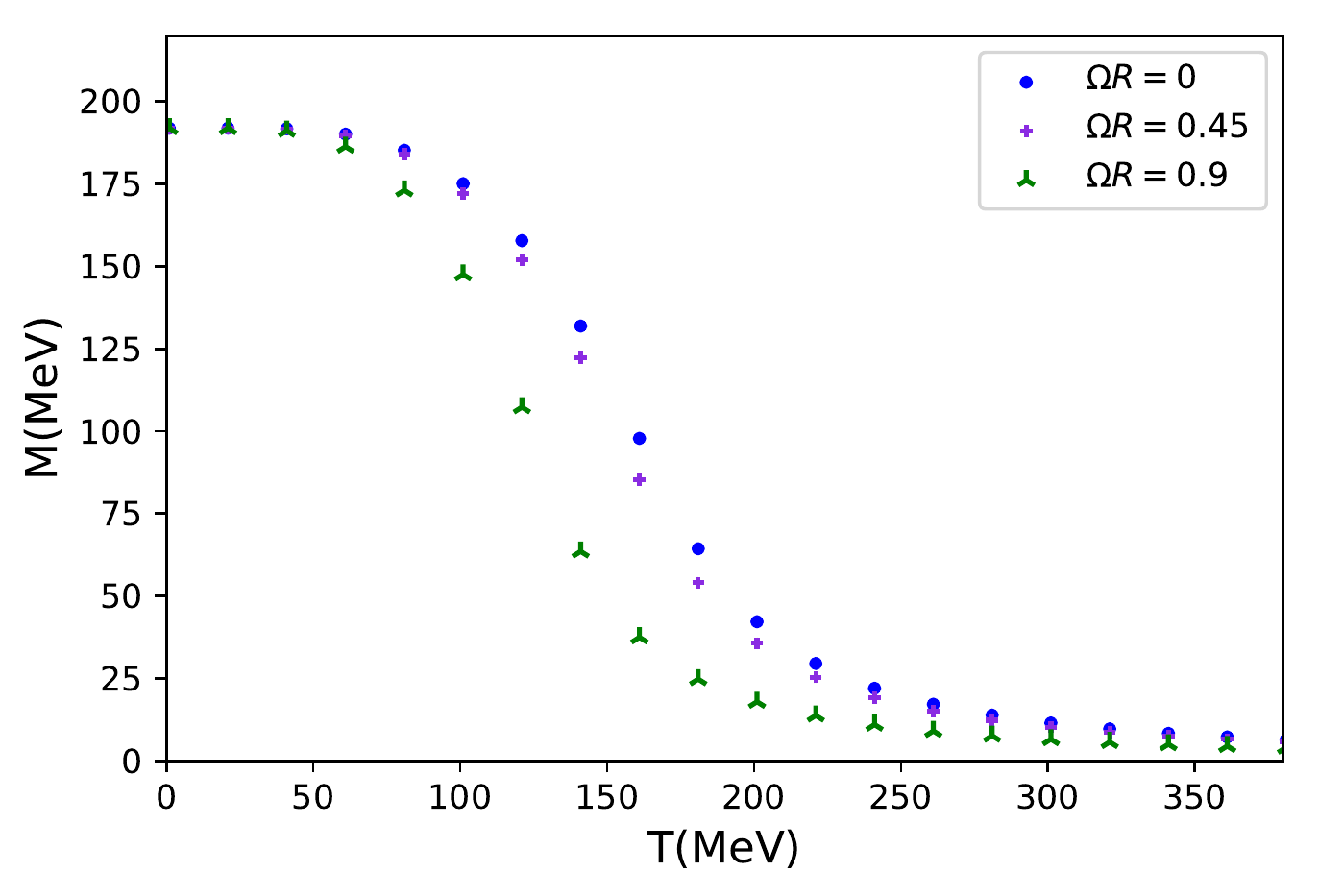} 
\end{minipage}
}
\centering
\caption{Effective mass as a function of $T$ at various $\Omega$.}
\label{fig1}
\end{figure*}

\section{Description in Rotating Frame} \label{II}
If we set the rotating axis to be the $z$ axis, the metric of a rotating frame with an angular velocity $\Omega$ can be written as \cite{cylinder}
\begin{equation}
g_{\mu \nu}=\left(\begin{array}{cccc}
{1-\left(x^{2}+y^{2}\right) \Omega^{2}} & {y \Omega} & {-x \Omega} & {0} \\
{y \Omega} & {-1} & {0} & {0} \\
{-x \Omega} & {0} & {-1} & {0} \\
{0} & {0} & {0} & {-1}
\end{array}\right).
\end{equation}
We adopt the convention that $\hat{i},\hat{j}\cdots=\hat{t},\hat{x},\hat{y},\hat{z}$ and $\mu,\nu\cdots=t,x,y,z$ refer to the Cartesian coordinate in the local rest frame and the general coordinate in the rotating frame, respectively.

The Dirac equation of a fermion with a mass $M$ in the curved spacetime is 
\begin{equation}
[i\gamma^\mu(\partial_\mu+\Gamma^\mu)-M]\psi=0,
\end{equation}
where 
\begin{equation}
\begin{aligned}
&\Gamma_\mu=-\frac{i}{4}\omega_{\mu \hat{i}\hat{j}}\sigma^{\hat{i}\hat{j}},\\
&\omega_{\mu \hat{i}\hat{j}}=g_{\alpha\beta}e^\alpha_{\hat{i}}(\partial_\mu e^{\beta}_{\hat{j}}+\Gamma^{\beta}_{\nu\mu} e^\mu_{\hat{j}}),\\
&\sigma^{\hat{i}\hat{j}}=\frac{i}{2}[\gamma^{\hat{i}},\gamma^{\hat{j}}],
\end{aligned}
\end{equation}
with the Christoffel connection, $\Gamma^\lambda_{\mu\nu}=\frac{1}{2}g^{\lambda\sigma}(g_{\sigma\nu,\mu}+g_{\mu\sigma.\nu}-g_{\mu\nu,\sigma})$, and the gamma matrix in curved space-time, $\gamma^\mu=e^\mu_{\hat{i}}\gamma^{\hat{i}}$. The vierbein $e^\mu_{\hat{i}}$ connects the general coordinate with the Cartesian coordinate in the rest frame, $x^\mu=e^\mu_{\hat{i}}x^{\hat{i}}$. Then the Dirac equation can be reduced to \cite{cylinder}
\begin{equation}
\label{Dirac}
\left[\gamma^{\hat{t}}\left(i \partial_{t}+\Omega J_{z}\right)+i \gamma^{\hat{x}} \partial_{x}+i \gamma^{\hat{y}} \partial_{y}+i \gamma^{\hat{z}} \partial_{z}-M\right] \psi=0.
\end{equation}
So, we can regard the Hamiltonian in a rotating frame has shifted as $\hat{H}\to \hat{H}-\Omega J_z$, where $\hat{H}$ is the Hamiltonian in the rest frame.

There are spherical wave solutions to the Dirac equation (\ref{Dirac}). We can check that a complete set of commutating operators consist of $\hat{H}, \hat{J}^2, \hat{J}_z, \hat{K}=\gamma^{\hat{t}}(\vec{\Sigma}\cdot\vec{J}-{\hbar}/{2})$, so the eigenstates can be labeled by a set of eigenvalues: energy $E$, total angular-momentum quantum number $j=1/2,3/2,\cdots$, total z-angular-momentum number $m_j=-j,-j+1,\cdots,j$, and $\kappa=\pm (j+1/2)$ \cite{Sakurai}. To bound the system in a sphere, we impose the MIT boundary condition \cite{MIT1,MIT2},
\begin{equation}\label{bc}
-i\hat{r}\cdot\vec{\gamma}\psi(t,r,\theta,\phi)|_{r=R}=\psi(t,r,\theta,\phi)|_{r=R},
\end{equation}
where $\hat{r}$ is the unit vector normal to the sphere surface, and $\vec{\gamma}=(\gamma^1,\gamma^2,\gamma^3)$. This boundary condition makes the normal component of the fermionic current $j^\mu=\overline{\psi}\gamma^\mu\psi$ to be zero on the surface. The causality requires $\Omega R<1$. With this boundary condition, the allowed values of momentum $p$ are given by the following eigen-equations \cite{Greiner}
\begin{equation}\label{engin}
j_{l_\kappa}(pR)=-\operatorname{sgn}(\kappa)\frac{p}{E_p+M}j_{\overline{l}_\kappa}(pR),
\end{equation}
where $$l_{\kappa}=\left\{\begin{array}{cl}{-\kappa-1} & {\text { for } \kappa<0} \\ {\kappa} & {\text { for } \kappa>0}\end{array}\right.,$$
$$\overline{l}_{\kappa}=\left\{\begin{array}{cc}{-\kappa} & {\text { for } \kappa<0} \\ {\kappa-1} & {\text { for } \kappa>0}\end{array}\right.,$$
and $j_l(x)$ is the $l$th ordered spherical Bessel function. We label the $i$th nonzero root of Eq. (\ref{engin}) with $j$ and $\kappa$ by $p_{j\kappa,i}$, then the corresponding energy with z-angular-momentum $m_j$ is $E=\sqrt{p_{j\kappa,i}^2+M^2}-\Omega m_j$. Once we obtain the spectrum in rotating frame, we can turn to the modification of the NJL model.

\section{Chiral phase transition in a Rotating Sphere}\label{III}
The Lagrangian of the two-flavor NJL model is 
\begin{equation}
\mathscr{L}=\overline{\psi}(i\gamma^\mu\partial_\mu-m_0)\psi+G[(\overline{\psi}\psi)^2+(\overline{\psi}i\gamma^5\tau\psi)^2],
\end{equation}
where $m_0$ is the current quark mass, and $G$ is the effective coupling. In the mean field approximation, the gap equation at finite temperature is given as \cite{NJLreview}
\begin{equation}
M=m_0+4GN_cN_f\int\frac{d^3p}{(2\pi)^3}\frac{M}{E_p}(1-\frac{2}{1+\mathrm{exp}(\frac{E_p}{T})}),
\end{equation}
where $N_c=3, N_f=2$ is the number of colors and flavors, respectively. To consider the finite size effects, we replace the integral over continuous momentum modes with a sum over discrete modes \cite{zzhang}. As for the rotation, we follow \cite{unbound1} to identify $\Omega m_j$ as a chemical potential, then the gap equation is modified to
\begin{equation}
\label{gap}
\begin{aligned}
M=&m_0+2GN_cN_f\frac{1}{V}\sum_{j}\sum_{p_{j\kappa,i}}\sum_{m_j=-j}^{m_j=j}\frac{M}{E_{p_{j\kappa,i}}}(1-\\ &\frac{1}{1+\mathrm{exp}(\frac{E_{p_{j\kappa,i}}-\Omega m_j}{T})}+\frac{1}{1+\mathrm{exp}(\frac{E_{p_{j\kappa,i}}+\Omega m_j}{T})}).
\end{aligned}
\end{equation}
where $E_{p_{j\kappa,i}}=\sqrt{p_{j\kappa,i}^2+M^2}$, and $V$ is the volume of the sphere. Here we note that in this paper we treat the chiral condensation $\left\langle\overline{\psi}\psi \right\rangle=(M-m_0)/(2G)$ to be homogeneous, while it should vary with coordinates in general. It is shown that the condensation is almost a constant inside a rest cylinder with NJL model in \cite{bound1}, so we think this homogeneous approximation is valid at least when the rotation speed is slow.
\begin{figure}[h]
\centering
\subfigure[ R=0.985 fm]{
\begin{minipage}{0.43\textwidth}
\centering
\includegraphics[width=1\linewidth]{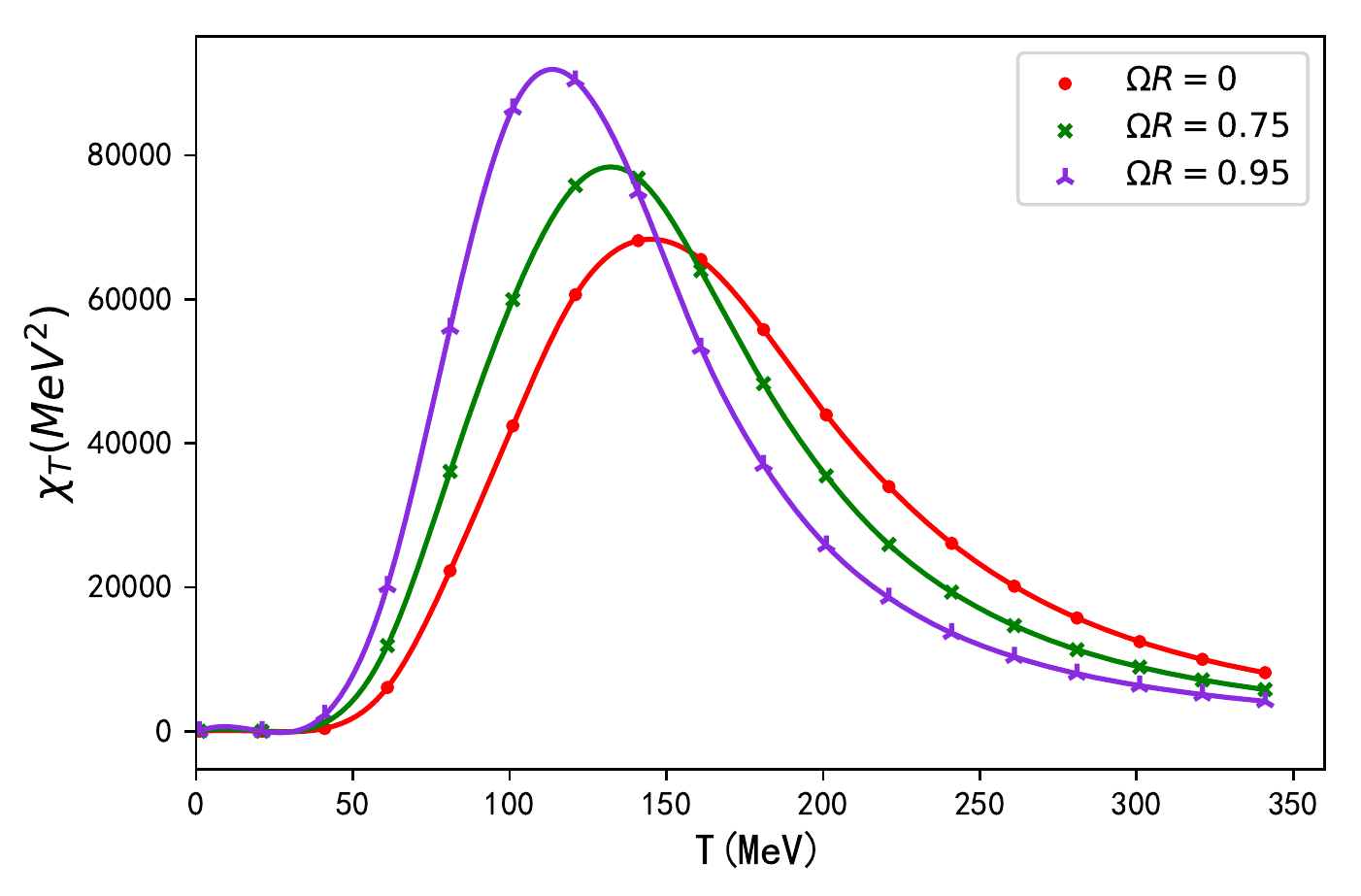} 
\end{minipage}
}
\subfigure[ R=2.955 fm]{
\begin{minipage}{0.43\textwidth}
\centering
\includegraphics[width=1\linewidth]{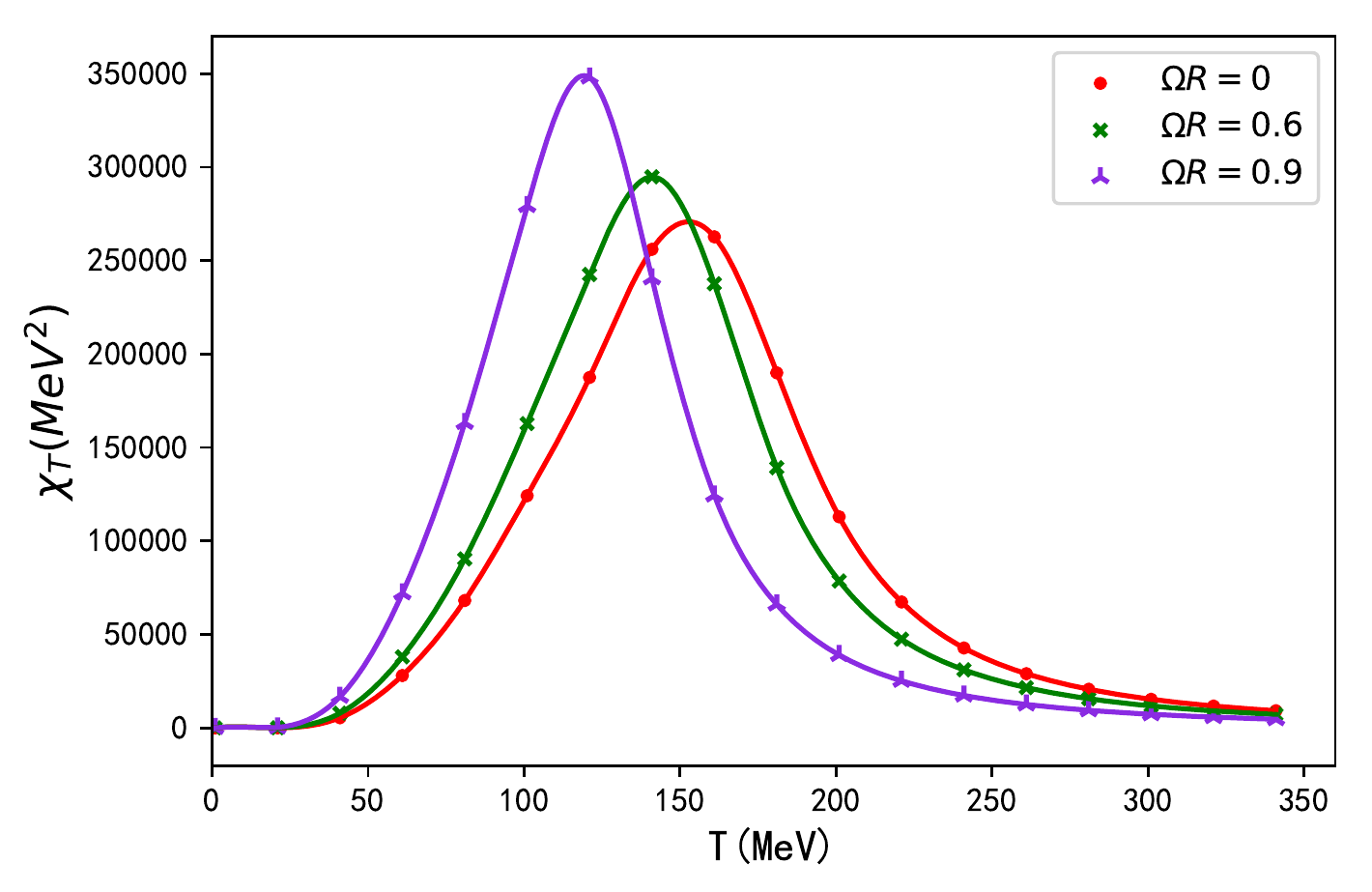} 
\end{minipage}
}
\caption{Thermal susceptibility as a function of $T$ at various $\Omega$.}
\label{fig2}
\end{figure}
Since NJL model is a nonrenormalizable model, regularization is needed. A usual adopted regularization is the three momentum cutoff which ignores the high-frequency modes. But in a finite volume system, only the high-frequency modes can exist, so here we adopt the proper time regularization~\cite{NJLreview}, which takes into account the contribution of all modes. The key equation of this regularization is a replacement
\begin{equation}
\frac{1}{A^{n}} \rightarrow \frac{1}{(n-1) !} \int_{\tau_{U V}}^{\infty} d \tau \tau^{n-1} e^{-\tau A}.
\end{equation}
where $\tau_{UV}$ is the regularization parameter. We enforce this replacement to the divergent term in the gap equation (\ref{gap}).

Before our numerical calculation, we do some qualitative analysis. First, we consider the zero temperature case; one can find the gap equation reduces to
\begin{equation}
\begin{aligned}
M=&m_0+2GN_cN_f\frac{1}{V}\sum_{j}\sum_{p_{j\kappa,i}}\sum_{m_j=-j}^{m_j=j}\frac{M}{E_{p_{j\kappa,i}}},
\end{aligned}
\end{equation}
which is independent of $\Omega$. So rotation has no effect at zero temperature, which has been already discussed in~\cite{cylinder,bound1}.

At a finite temperature, one can show for any $\Omega$ that
\begin{equation}
\begin{aligned}
&\sum_{m_j=-j}^{m_j=j}(1-\frac{1}{1+\mathrm{exp}(\frac{E_{p_{j\kappa,i}}-\Omega m_j}{T})}+\frac{1}{1+\mathrm{exp}(\frac{E_{p_{j\kappa,i}}+\Omega m_j}{T})})\\
&\leq \sum_{m_j=-j}^{m_j=j}(1-\frac{1}{1+\mathrm{exp}(\frac{E_{p_{j\kappa,i}}}{T})}+\frac{1}{1+\mathrm{exp}(\frac{E_{p_{j\kappa,i}}}{T})}).
\end{aligned}
\end{equation}
Thus rotation will lead to a smaller effective mass at a finite temperature, which is consistent with what we explained in the Introduction.

Now let us study how the effective mass changes with $T$ and $\Omega$. In Fig. \ref{fig1}, we show $M$ as a function of $T$ for various $\Omega$ at some fixed radii. The parameters are chosen as $m_0=5 \ \mathrm{MeV}, \tau_{UV}=1/1080\ \mathrm{MeV}^{-2}, G=3.26\times 10^{-6} \ \mathrm{MeV}^{-2}$.

We can observe that the effective mass decreases with decreasing the volume, as we explained in the Introduction and found in a previous study~\cite{zzhang}. At a fixed radius, the rotation suppresses the chiral condensation at a finite temperature, thus making the phase conversion happen at a lower temperature. However, it has no effect at zero temperature. One can observe that rotation has considerable effects only when $\Omega R$ are close to unity, which can be achieved in HIC experiments. Also, one can observe that finite size has stronger effects on the phase conversion than the rotation for a system in a fetermeter scale.
To see how rotation influences a phase conversion more clearly, we present the thermal susceptibility $\chi_T=\frac{\partial\left\langle\overline{\psi}\psi \right\rangle}{\partial T}$ as a function of temperature for various $\Omega$ at two different radii in Fig. \ref{fig2}. We can observe that the peaks occur at a lower temperature when $\Omega$ increases, which indicates the phase conversion happens at a lower temperature. Also, the peaks becomes shaper when $\Omega$ increases.

\section{Summary}\label{IV}
We studied the chiral phase transition with the NJL model in a rotating sphere. It was found that both rotation and finite size effects catalyze the phase conversion from the hadron phase to the quark phase, but the latter has stronger effects than the former. Furthermore, rotation can induce nontrivial transport phenomenon, which is very important for exploring the properties of hot and dense QCD matter created in the HIC experiments. On the other hand, the analogy and interplay between rotation and a magnetic field has been discussed in other literatures~\cite{unbound1}; it is also a topic that deserves discussion since HIC experiments produce a strong magnetic field as well. Our work may be helpful for studying these problems in a more realistic way in the future.

\section*{Acknowledgements}
This work is supported in part by the National Natural Science Foundation of China (under Grants No. 11475085, No. 11535005, No. 11905104, and No. 11690030) and by Nation Major State Basic Research and Development of China (2016YFE0129300). X. Luo is supported by the National Key Research and Development Program of China (2018YFE0205201),  the National Natural Science Foundation of China (Grants No. 11828501, No. 11575069, No. 11890711 and No. 11861131009). 

\bibliography{ref}
\end{document}